\documentclass[
 aip,
 jmp,
 amsmath,
 amssymb,
 reprint, 12pt, onecolumn
]{revtex4-1}

\usepackage{braket}
\usepackage{graphicx}
\usepackage{dcolumn}
\usepackage{bm}
\usepackage{array}
\usepackage{booktabs}
\usepackage{multirow}
\usepackage{rotating}
\usepackage{amsmath}
\usepackage{appendix}
\usepackage{color}
\usepackage{natbib}
\usepackage{makeidx}

\begin{document}

\title[]{Excited state geometry optimization with the density matrix renormalization group as applied to polyenes}
\author{Weifeng Hu}
\author{Garnet Kin-Lic Chan}
\affiliation{Department of Chemistry, Princeton University, Princeton, New Jersey 08544, USA}

\begin{abstract}
We describe and extend the formalism of 
state-specific analytic density matrix renormalization group (DMRG) 
energy gradients, first used by Liu et al (J. Chem. Theor. Comput. \textbf{9}, 4462 (2013)). 
We introduce a DMRG wavefunction maximum overlap following technique 
to facilitate state-specific DMRG excited state optimization.
Using DMRG configuration interaction (DMRG-CI) gradients 
we relax the low-lying singlet states of a series of \textit{trans}-polyenes up to C$_{20}$H$_{22}$. 
Using the relaxed excited state geometries as well as correlation functions, 
we elucidate the exciton, soliton, and bimagnon (``single-fission'') 
character of the excited states, and find evidence for a planar conical intersection.
\end{abstract}

\maketitle

\section{Introduction}\label{sec:intro}
The density matrix renormalization group\cite{White1999,Mitrushenkov2001, Chan2002, Chan2004, Zgid2009, Reiher2010review, Chan2011},
introduced by White~\cite{White1999}, has made large active space multireference quantum chemistry calculations routine.
In the chemistry context, there have been many improvements to White's algorithm, 
including orbital optimization\cite{Zgid2008, Zgid2008b, Ghosh2008, Luo2010}, spin-adaptation\cite{Zgid2008a, Sharma2012}, 
dynamic correlation treatments\cite{Neuscamman2010, Yanai2010, Kurashige2011}, 
and response theories\cite{DMRG_response_theory.Jon, DMRG_response_theory.Naoki}, to name a few. 
The DMRG has been applied in many different electronic structure problems, 
ranging from benchmark exact solutions of the Schr\"odinger equation 
for small molecules\cite{Legeza2003, Chan2003, Chan2004a, Sharma2014near, Sharma2014, Kurashige2011}, 
to active space studies of conjugated $\pi$-electron
systems\cite{Chan2005, Hachmann2006, Hachmann2007, Mizukami2012}, 
the elucidation of the ground- and excited states of multi-center 
transition-metal clusters\cite{Reiher2008, Kurashige2013, Sharma2014low, Wouters2014b, Kurashige2009, Fertitta2014}, 
computation of high-order correlation contributions to the binding energy of molecular crystals\cite{Yang2014}, 
relativistic calculations~\cite{knecht2014communication}, and 
the study of curve crossings in photochemistry\cite{Liu2013multireference}.

Energy gradients are crucial to electronic structure as they define equilibrium structures,
transition states, and reaction trajectories.
Analytic energy gradients, introduced by Pulay~\cite{pulay1969ab, pulay1977direct, pulay1979efficient, pulay1979systematic}, 
are preferred to numerical gradients, due to their low cost and numerical stability, and 
are now implemented for many single- and multi-reference 
quantum chemistry methods~\cite{CPKS,gradient_bible,CI_gradient,MRCI_gradient.1, MRCI_gradient.2,MRCC_gradient.1, MRCC_gradient.2}.

Analytic DMRG energy gradients were first used by Liu et al.~\cite{Liu2013multireference} 
in a study of the photochromic ring opening of spiropyran.
The theory, although simple, was not fully discussed. A contribution
of the current work is to provide a more complete exposition of the theory behind DMRG gradients.
A second contribution is to discuss their practical implemention 
in excited state geometry optimization. The simplest formulation of the gradients 
arises when the excited states are treated in a state-specific manner (that is,
 without orthogonality constraints to lower states). However, such DMRG calculations 
can be susceptible to root-flipping, for example, near conical intersections. 
Furthermore, DMRG wavefunctions are specified both by a choice of active space
as well as by discrete sets of quantum numbers associated with
each orbital (used to enforce global symmetries, such as the total
particle number). During an energy minimization, it is important that the wavefunction
changes smoothly.
Here, we present a state-specific DMRG analytic gradient algorithm that uses
a maximum overlap technique both to stably converge excited states, and to ensure
adiabatic changes of both the orbitals and the DMRG wavefunction during geometry changes.

Trans-polyenes are well-known examples of molecules with interesting ground- and excited state 
structure, and form the central motifs for a large set of of biological compounds, such as 
retinals and the carotenoids. 
Many calculations using semi-empirical models (such as the Pariser-Parr-Pople (PPP) model) as well as with {\it ab-initio} methods such as multi-reference self-consistent-field (MC-SCF), have been carried out to identify the low-lying electronic and geometric 
features~\cite{su1979solitons, su1980soliton, subbaswamy1981bond, boudreaux1983solitons, soos1983spin, aoyagi1985mcscf, Hayden1986, tavan1987electronic, bredas1990theoretical, serrano1993towards, Hirao1996, Shuai1997, nakayama1998theoretical, Fano1998, Yaron1998, Barford1997, Barford1998, Barford1999, Barford2001, Barford2002, Barford2002excitons, Barford2002breakdown, Barford2003, Ma2004, Ma2005, Ma2005theoretical, Ma2006, Ma2008, Ma2009, Ma2010}. 
These studies, in general, give the following qualitative picture for long even polyenes: 
1) Excitations, coupled to lattice relaxation, break the dimerization of the ground state and 
 lead to local geometrical defects~\cite{barford_book}. 
For example, adiabatic relaxation 
gives rise to a central polaronic feature
in the bright 1$^{1}$B$_{u}$ state~\cite{Barford1999, Barford2002breakdown, Ma2008, Ma2009}, 
as well as two-soliton or four-soliton structures for the dark 2$^{1}$A$_{g}$ state~\cite{Hayden1986, bredas1990theoretical, Barford1999}. 
2) For short polyenes ranging from ethylene to octatriene, studies of low-lying excited state relaxation pathways  suggest 
that non-planar molecular conformations are important at energy crossings~\cite{dormans1987dynamical, olivucci1993conical, celani1995excited, ito1997nonadiabatic, ben1998ab, brink1998triplet, krawczyk200011bu, garavelli2000relaxation, ostojic2001ab, garavelli2001product, sampedro2002structure, dou2003detailed, nonnenberg2003restricted, kouppel2007multimode, sampedro2002structure, levine2008optimizing, Martinez2009, Liu2013}. 
This opens up the question of the nature of energy crossings and their associated geometries in the excited states of longer polyenes.
Here, using an {\it ab-initio} Hamiltonian and DMRG-CI analytic energy gradients, we revisit these questions
in the excited states of relatively long \textit{trans}-polyacetylenes, aiming for a more quantitative picture.

In Sec.~\ref{sec:general_grad} and \ref{sec:dmrg}, we start by reviewing 
analytic energy gradient theory and DMRG theory. In Sec.~\ref{sec:grad_dmrg} we discuss 
the formulation of DMRG analytic gradients.
In Sec.~\ref{sec:maxov}, we discuss the DMRG maximum 
overlap method for stable state-specific excited state calculations 
without orthogonality constraints.
In Sec.~\ref{sec:polyene} we apply our DMRG gradient algorithm to 
characterize the geometry and nature of the low-lying singlet 
states of the trans-polyenes.

\section{General energy gradient theory for variational methods}\label{sec:general_grad}

For completeness, we briefly recall analytic energy gradient theory for variational wavefunctions. 
The energy referred to is the Born-Oppenheimer potential, the
sum of the electronic energy and the nuclear-nuclear repulsion. 
The nuclear-nuclear repulsion gradient is trivial. The electronic energy $\langle \Psi|\hat{H}|\Psi\rangle$ 
is the expectation value of the electronic Hamiltonian $\hat{H}$,
\begin{equation}\label{eq:qc_h}
 \hat{H} = \sum_{ij\sigma} t_{ij} c^{\dag}_{i\sigma} c_{j\sigma} + \frac{1}{2} \sum_{ijkl \sigma \sigma\prime} v_{ijkl} c^{\dag}_{i\sigma} c^{\dag}_{j\sigma\prime} c_{k\sigma\prime} c_{l\sigma},
\end{equation}
where $i,j,k,l$ are orthogonal (for example, molecular) orbital indices, and $\sigma, \sigma'= \{\uparrow,\downarrow \}$.
$\hat{H}$ depends on the orbital functions through the integrals
 $t_{ij}$ and $v_{ijkl}$. 
In a typical implementation, these orthogonal orbitals
are represented as a linear combination of atomic orbitals (AO's) 
with orbital coefficients \textbf{C}: $\ket{i}=\sum_\mu C_{i\mu} \ket{\mu}$, 
where we use Greek letters ($\mu, \nu$) to denote AO orbitals. 
The geometry dependence of the integrals arises from the AO functions, 
which (as Gaussian type basis functions) explicitly depend on the nuclear positions, 
as well as through the LCAO coefficient matrix $\textbf{C}$, which also changes with geometry. 
The Hamiltonian may be regarded as a function of the nuclear coordinates $\{ a_i\}$ and
the coefficient matrix $\mathbf{C}$: $\hat{H}(\{ a_i\}, \mathbf{C})$.
As the electronic wavefunction $\ket{\Psi}$ depends on 
variational parameters $\{ c_i\}$, the electronic energy 
is a function of $\{ a_{i} \}$, $\mathbf{C}$, and $\{c_i\}$: 
$E(\{a_{i}\}, \textbf{C}, \{c_{i}\} )$.
The energy gradient with respect to nuclear coordinate $a$  takes the form
\begin{equation}\label{eq:grad_general.1}
\begin{aligned}
\dfrac{\textit{d} E}{ \textit{d} a} = 
            \dfrac{\partial E}{ \partial a} + 
            \sum_{i\mu}\dfrac{\partial E}{\partial C_{i\mu}} \dfrac{ \textit{d} C_{i\mu}}{ \textit{d} a} + 
            \sum_{i}\dfrac{ \partial E }{\partial c_{i}} \dfrac{ \textit{d} c_{i}}{ \textit{d} a}.
\end{aligned}
\end{equation}
If $\ket{\Psi}$ is determined variationally, the energy is stationary to changes of $\{c_{i}\}$, thus the third term vanishes. 
The gradient then only depends on the change in nuclear coordinates and orbital coefficients,
\begin{equation}\label{eq:grad_general.2}
\begin{aligned}
\dfrac{\textit{d} E}{ \textit{d} a} = 
            \dfrac{\partial E}{ \partial a} + 
            \sum_{i}\dfrac{\partial E}{\partial C_{i}} \dfrac{ \textit{d} C_{i}}{ \textit{d} a}.
\end{aligned}
\end{equation}

It is convenient to rewrite the energy gradient in terms of density matrices. 
The energy is expressed as
\begin{equation}\label{eq:eng_pdm}
 E =  \sum_{ij} t_{ij} \gamma_{ij} +
     \sum_{ijkl} v_{ijkl}  \Gamma_{ijkl},
\end{equation}
where $\gamma_{ij}=\sum_{\sigma}\bra {\Psi}  c^{\dag}_{i\sigma} c_{j\sigma} \ket{ \Psi}$ 
and $\Gamma_{ijkl}=\frac{1}{2} \sum_{\sigma\sigma'}\bra{ \Psi}  c^{\dag}_{i\sigma} c^{\dag}_{j\sigma\prime} c_{k\sigma\prime} c_{l\sigma} \ket{ \Psi}$ are the
one- and two-particle density matrices. 
Since the second-quantized operators have no dependence on geometry, 
and the wavefunction depends only on $\{c_i\}$, it follows from 
Eq.~(\ref{eq:grad_general.2}) that the energy gradient
is expressed in terms of the one- and two-electron derivative integrals and density matrices,
\begin{equation}\label{eq:grad_pdm}
\begin{aligned}
 \frac{dE}{da} = \sum_{ij}\frac{dt_{ij}}{da} \gamma_{ij} + \sum_{ijkl}\frac{d v_{ijkl}}{da} \Gamma_{ijkl}.
\end{aligned}
\end{equation}
The one- and two-electron derivative integrals involve the orbital derivative (response), $d \mathbf{C}/da$.
Writing
\begin{equation}\label{eq:mo_drv}
\frac{d C_{i \mu}}{da } = \sum_{j} U^{a}_{ij} C_{j \mu} 
\end{equation}
gives
\begin{equation}\label{eq:eng_grad_mo}
\begin{aligned}
\frac{dE}{da} = &\sum_{ij}\frac{\partial h_{ij}}{\partial a }\gamma_{ij} + \sum_{ijkl} \frac{\partial v_{ijkl}}{\partial a} \Gamma_{ijkl}-\sum_{ij}X_{ij}\frac{\partial S_{ij}}{\partial a}\\
                & + \sum_{ij}U^{a}_{ij}(X_{ij}-X_{ji}),
\end{aligned}
\end{equation}
where $S_{ij}=\langle i|j\rangle$ is the overlap matrix of the orthogonal orbitals $i$ and $j$, 
\begin{align}
S_{ij} = \sum_{\mu \nu} C_{i\mu} S_{\mu \nu} C_{j\nu},
\end{align}
and $S_{\mu \nu}=\langle \mu|\nu\rangle$ is the overlap matrix in the underlying AO basis.
 $\mathbf{X}$  is the Lagrangian matrix in the Generalized Brillouin Theorem (GBT)~\cite{GBT}, given as
\begin{equation}\label{eq:x_mat}
X_{ij} = \sum_{m} h_{im} \gamma_{mj} + 2 \sum_{mkl} v_{imkl} \Gamma_{jmkl},
\end{equation}
characterizing the energy cost of electronic excitation from $i$th orbital to $j$th orbital.
The gradient formula can be rewritten entirely in terms of AO quantities~\cite{gradient_bible},
\begin{equation}\label{eq:eng_grad_ao}
\begin{aligned}
&\frac{d E}{ d a} = \sum_{\mu\nu} \gamma_{\mu\nu} \frac{d h_{\mu\nu}}{\partial a} + \sum_{\mu\nu\rho\sigma} \Gamma_{\mu\nu\rho\sigma} \frac{d v_{\mu\nu\rho\sigma}}{\partial a} \\
&- 2 \sum_{\mu\nu} \sum_{i>j}(1-\frac{\delta_{ij}}{2})C_{\mu}^{i}C_{\nu}^{j}X_{ji} \frac{d S_{\mu\nu}}{\partial a} + 2 \sum_{i>j} U^{a}_{ij}(X_{ij}-X_{ji})
\end{aligned}
\end{equation}
where $h_{\mu\nu}$, $\gamma_{\mu \nu}$, $v_{\mu\nu\rho\sigma}$, $\Gamma_{\mu \nu \rho \sigma}$ are the one- and two- particle integrals and reduced density matrices, respectively, in the AO basis.

In general, the orbital derivative $dC_{i\mu}/da$ requires the solution
of equations which couple the wavefunction coefficients $c_i$ to the orbital 
coefficients $C_{i\mu}$. 
However, there are two common situations where the orbital response is simplified.
The first is when the orbitals are defined independently 
of the correlated wavefunction, for example,
for Hartree-Fock (HF) or Kohn-Sham (KS) orbitals. 
Using the Hartree-Fock canonical orbitals as an example, 
the orbital response $\textbf{U}^{a}$ is defined by the Hartree-Fock
convergence condition, 
\begin{equation}\label{eq:cphf.1}
\frac{d F_{ij}}{da} = 0 (i\ne j).
\end{equation}
which leads to the definition of the $\textbf{U}^{a}$ matrix
\begin{equation}\label{eq:cphf.2}
U^{a}_{ij} = \frac{1}{(\epsilon_{j} - \epsilon_{i})}(\sum_{k}^{vir} \sum_{l}^{d.o.} A_{ij, ai} U^{a}_{kl} + B^{a}_{ij}),
\end{equation}
where
\begin{equation}\label{eq:cphf.3}
\begin{aligned}
&A_{ij,kl} = 4 v_{ijkl} - v_{ikjl} - v_{iljk}\\
&B^{a}_{ij} = F^{a}_{ij} - S^{a}_{ij}\epsilon_{j} - \sum_{jk} S^{a}_{kl}(2 v_{ijkl}-v_{ikjl}),
\end{aligned}
\end{equation}
with $F^{a}_{ij}=\partial F_{ij}/\partial a$, $S^a_{ij}=\partial S_{ij}/\partial a$, 
and the various $\epsilon$ are HF orbital energies. 
Eq.~(\ref{eq:cphf.2}) is the coupled-perturbed Hartree-Fock (CPHF) equation\cite{CPHF}, 
and uniquely defines the $\textbf{U}^{a}$ matrix elements for canonical orbitals. 
In a similar way, other types of orbital response, for example for the Kohn-Sham orbitals,
or localized Hartree-Fock orbitals, can be computed from the corresponding 
coupled-perturbed single-particle equations~\cite{CPKS,CP-L}.

The second simplifying case is when the correlated wavefunction energy is itself
stationary with respect to orbital variations. In this case 
$X_{ij}-X_{ji}=0$, and the orbital response is not required, even though 
it is formally coupled to the correlated variational wavefunction coefficients. 
The energy gradient reduces to the simpler form,
\begin{equation}\label{eq:dmrgscf_grad_ao}
\begin{aligned}
&\frac{d E}{ d a} = \sum_{\mu\nu} \gamma_{\mu\nu} \frac{d h_{\mu\nu}}{\partial a} + \sum_{\mu\nu\rho\sigma} \Gamma_{\mu\nu\rho\sigma} \frac{d v_{\mu\nu\rho\sigma}}{\partial a} \\
&- 2 \sum_{\mu\nu} \sum_{i>j}(1-\frac{\delta_{ij}}{2})C_{\mu}^{i}C_{\nu}^{j}X_{ji} \frac{d S_{\mu\nu}}{\partial a} + 2 \sum_{i>j} U^{a}_{ij}(X_{ij}-X_{ji}).
\end{aligned}
\end{equation}

\section{General DMRG Theory}\label{sec:dmrg}

The DMRG is a variational wavefunction method\cite{DMRG_Lagrangian}.
For a set of $L$ orthogonal orbitals (where the states for the $i$th orbital are $\ket{\sigma_i}=\left\{\ket{0},\ket{\uparrow} ,\ket{\downarrow} ,\ket{\uparrow \downarrow}  \right\} $) we
choose a partitioning of the orbitals into a  {left block}, {single site}, and {right block}, consisting of orbitals $\{1...l-1\}$, $\{l\}$ and $\{l+1...L\}$, respectively.
The corresponding canonical ``one-site'' DMRG wavefunction takes the matrix product form
\begin{equation}\label{eq:dmrg_wf_mps}
\begin{aligned}
\ket{ \Psi} = & 
    \sum_{\sigma_{1} \sigma_{2} ... \sigma_{L} } 
        \mathbf{L}^{\sigma_{1}} \mathbf{L}^{\sigma_{2}} ... \mathbf{L}^{\sigma_{l-1}} \\
                    & \times \mathbf{C}^{\sigma_{l}}  
                            \mathbf{R}^{\sigma_{l+1}} \mathbf{R}^{\sigma_{l+2}}... \mathbf{R}^{\sigma_{L}}
                                    \ket{ \sigma_{1} \sigma_{2} ... \sigma_{L}}. \\
\end{aligned}
\end{equation}
The (rotation) matrices $\mathbf{L}^{\sigma_i}$ and $\mathbf{R}^{\sigma_i}$ are of dimension $M \times M$, except for the first and last which 
are of dimension $1 \times M$ and $M \times 1$ respectively. They satisfy the left- and right-canonical conditions
\begin{align}
\sum_{\sigma_i} {\mathbf{L}^{\sigma_i}}^T \mathbf{L}^{\sigma_i} &= \mathbf{1} \notag\\
\sum_{\sigma_i} {\mathbf{R}^{\sigma_i}} {\mathbf{R}^{\sigma_i}}^T &= \mathbf{1} \label{eq:lrcanonical}
\end{align}
while the $\mathbf{C}^{\sigma_l}$ (wavefunction) matrix satisfies the normalization condition
\begin{align}
\mathrm{tr} \sum_{\sigma_l} {\mathbf{C}^{\sigma_l}}^T \mathbf{C}^{\sigma_l} = 1. \label{eq:ccanonical}
\end{align}
Together, $\{\mathbf{L}^{\sigma_i}\}$, $\{\mathbf{C}^{\sigma_l}\}$ and $\{\mathbf{R}^{\sigma_i}\}$ contain the variational 
parameters. As in other variational methods, the coefficients of the matrices
are determined by minimizing the energy. In principle, a direct gradient minimization
of the energy with respect to all the matrices, subject to the canonical
conditions Eq.~(\ref{eq:lrcanonical}),~(\ref{eq:ccanonical}), may be performed. In practice,
the DMRG sweep algorithm is normally used. Here, at a given step $l$ of the sweep,
corresponding to the block partitioning $\{1...l-1\}$, $\{l\}$ and $\{l+1...L\}$,
the energy is minimized only with respect to the $\mathbf{C}^{\sigma_l}$ wavefunction matrix,
with the $\{\mathbf{L}^{\sigma_i}\}$,  $\{\mathbf{R}^{\sigma_i}\}$ rotation matrices held fixed.
The minimizing $\mathbf{C}^{\sigma_l}$ is obtained from an effective ground-state eigenvalue problem
\begin{align}
\mathbf{H} \mathbf{c}=E \mathbf{c}
\end{align}
where $c$ denotes $\mathbf{C}^{\sigma_l}$ flattened into a single vector, and $\mathbf{H}$
denotes $\hat{H}$ expressed in the basis of renormalized basis states defined by the $\{\mathbf{L}^{\sigma_i}\}$,  $\{\mathbf{R}^{\sigma_i}\}$ matrices~\cite{DMRG_Lagrangian}. In the next step of the sweep, the single site is moved from $l$ to $l+1$ (or $l$ to $l-1$ in a backwards sweep). 
To satisfy the new canonical form with the single site at $l+1$,
where the $\mathbf{C}^{\sigma_l}$ matrix is 
replaced by an $\mathbf{L}^{\sigma_l}$ matrix, 
and the $l+1$ site
is associated with a new $\mathbf{C}^{\sigma_{l+1}}$ matrix,
we use the gauge relations,
\begin{align}
\mathbf{C}^{\sigma_l} &= \mathbf{L}^{\sigma_l} \mathbf{\Lambda}\notag\\
\mathbf{C}^{\sigma_{l+1}}& = \mathbf{\Lambda} \mathbf{R}^{\sigma_{l+1}} \label{eq:qr_decomp}.
\end{align}
By sweeping through all the partitions $l=1\ldots L$, and minimizing
with respect to the $\mathbf{C}^{\sigma_l}$ matrix at each partition, the DMRG sweep algorithm 
ensures that all the variational degrees of freedom in the DMRG wavefunction are optimized.

An important aspect of DMRG calculations is the enforcement of symmetries, including global
symmetries such as the total particle number and spin. In the DMRG wavefunction, Abelian global symmetries (such as
total particle number) are enforced by local quantum numbers. For example, to enforce a
total particle number of $N$ in the wavefunction,  
each value of the 3 indices $\sigma$, $i, j$ in the matrix elements $\mathbf{L}^{\sigma}_{ij}$, $\mathbf{R}^{\sigma}_{ij}$, $\mathbf{C}^{\sigma}_{ij}$ can be associated with an
additional integer $N_i$, $N_j$, $N_\sigma$. (These values can be interpreted in terms
of the particle numbers of the renormalized states (for $N_i$ and $N_j$) and for the states of the single site (for $N_\sigma$)). Then, a total particle number of $N$ is
enforced with the rules:
\begin{align}
\mathbf{L}: N_i+N_\sigma &= N_j \notag\\
\mathbf{R}: N_j+N_\sigma &= N_i \notag\\
\mathbf{C}: N_i+N_\sigma+N_j &= N.
\end{align}
Applying these conditions to $\mathbf{L}^{\sigma}_{ij}$, $\mathbf{R}^{\sigma}_{ij}$, $\mathbf{C}^{\sigma}_{ij}$ means that the matrices have a block-sparse structure, which is important to maintain during geometry optimization.

\section{State-specific DMRG analytic energy gradients}\label{sec:grad_dmrg}

At convergence of the above (one-site) DMRG sweep algorithm, the contribution
of the wavefunction coefficients to the 
gradient ($d c_i/d a$ in Eq.~(\ref{eq:grad_general.1})) vanishes, as expected 
for a variational wavefunction method. Thus the analytic energy 
gradient theory for variational wavefunctions described in Sec.~\ref{sec:general_grad} can be applied. 

We will consider energy gradients for two kinds of DMRG calculations. 
The first are DMRG configuration interaction (DMRG-CI) analytic gradients, using 
HF canonical orbitals. In this case, the orbital response is given by 
the CPHF equations, presented in Sec.~\ref{sec:general_grad}. 
The DMRG calculations are carried out within an active space, chosen as 
a subset of the canonical orbitals. Because the DMRG wavefunction is not 
invariant to rotations of the active space orbitals for small $M$,
the contribution of the active orbital response must be computed specifying a particular
orbital choice (rather than just their manifold), such as the canonical HF orbitals. 

The algorithm to compute the DMRG-CI analytic gradient with HF canonical orbitals is as follows:
\begin{enumerate}\label{enum:dmrgci_grad_step}
 \item \label{enum:dmrgci_grad_step.1} Solve the HF equations for 
the canonical orbital coefficient matrix \textbf{C}.
 \item \label{enum:dmrgci_grad_step.2} Select an active space, and solve for the DMRG wavefunction 
in this space. Compute the one- and two-particle reduced density 
matrices $\gamma_{ij}$ and $\Gamma_{ijkl}$ at the convergence of the single-site sweep algorithm.
 \item \label{enum:dmrgci_grad_step.3} Compute the AO derivative 
integrals ${d h_{\mu\nu}}{/d a}$, ${d v_{\mu\nu\rho\sigma}}/{ d a}$ and ${d S_{\mu\nu}}/{d a}$, 
and the \textbf{X} matrix in Eq.~(\ref{eq:x_mat}).
 \item \label{enum:dmrgci_grad_step.4} Use the derivative integrals to construct the 
CPHF equation in Eq.~(\ref{eq:cphf.2}) 
(or the equivalent $Z$-vector equation\cite{gradient_bible}) and 
solve for $\textbf{U}^{a}$ for all nuclear coordinates.
 \item \label{enum:dmrgci_grad_step.5} Compute the energy gradient by 
contractions of all the above integrals and matrices 
according to Eq.~(\ref{eq:eng_grad_mo}) or (\ref{eq:eng_grad_ao}).
\end{enumerate}

The second kind of DMRG calculation we consider is a DMRG 
complete active space self-consistent field (DMRG-CASSCF) calculation.
For DMRG-CASSCF wavefunctions, the DMRG energy is stationary to any orbital rotation, thus
\begin{equation}\label{eq:gbt}
X_{ij}-X_{ji} = 0
\end{equation}
and by Eq.~(\ref{eq:eng_grad_mo}) and (\ref{eq:eng_grad_ao}) this means 
that the orbital response is not required even though it is coupled to the 
response of the DMRG wavefunction. However,
because the DMRG wavefunction is not invariant to active space rotations for small $M$, 
it is necessary to optimize the active-active rotations also, unlike in a traditional
CASSCF calculation. Alternatively,
if active-active rotations are omitted, the DMRG-CASSCF gradient can be viewed as an 
approximate gradient with a controllable error from active-active 
contributions (which vanishes as $M$ is extrapolated to $\infty$.)

The algorithm for the DMRG-CASSCF gradient is:
\begin{enumerate}\label{enum:dmrgscf_grad_step}
 \item \label{enum:dmrgscf_grad_step.1} Solve for the DMRG-CASSCF orbitals 
with the {one-site} DMRG wavefunction. In each macroiteration:
   \begin{enumerate}
     \item \label{enum:dmrgscf_grad_step.1.1} Solve for the one-site state-specific 
DMRG wavefunction, and compute the one- and two-body 
reduced density matrices $\gamma_{ij}$ and $\Gamma_{ijkl}$.
     \item \label{enum:dmrgscf_grad_step.1.2} Using $\gamma_{ij}$ and $\Gamma_{ijkl}$, compute 
the orbital gradient and Hessian, both of which include elements for active-active rotations.
     \item \label{enum:dmrgscf_grad_step.1.3} Update the orbitals with the orbital rotation matrix.
   \end{enumerate}
 \item \label{enum:dmrgscf_grad_step.2} Compute the AO density matrices 
$\gamma_{\mu\nu}$ and $\Gamma_{\mu\nu\rho\sigma}$ at the convergence of DMRG-CASSCF.
 \item \label{enum:dmrgscf_grad_step.3} Compute the AO derivative 
integrals ${d h_{\mu\nu}}/{d A}$, ${d (\mu\nu|\rho\sigma)}/{ d A}$ and ${d S_{\mu\nu}}/{d A}$.
 \item \label{enum:dmrgscf_grad_step.4} Contract all the above integrals 
and matrices using Eq.~(\ref{eq:dmrgscf_grad_ao}) to obtain the energy gradients.
\end{enumerate}

\section{Adiabatic orbital and wavefunction propagation and excited state tracking}\label{sec:maxov}

Geometry optimization requires adiabatically propagating along a potential energy surface.
For a DMRG calculation, this means that in each geometry step, the orbitals defining
the active space should change continuously, and the quantum numbers 
and associated block-sparsity pattern of the matrices should not change. 
The former can be achieved using maximum overlap techniques, while the 
latter can be done by fixing the quantum numbers at the initial geometry.
For state-specific excited state calculations, the maximum overlap technique 
is further important to prevent root-flipping.
Root flipping in state-specific DMRG calculations arises because the matrices 
optimized in the wavefunction for one state (\ref{eq:dmrg_wf_mps}) are not 
optimal for another state~\cite{Dorando2007}.
(Note that the gradient formalism presented above
is only valid for state-specific, rather than state-averaged, DMRG calculations).

\subsection{Orbital maximum overlap}\label{subsec:maxov_mo}

The maximum overlap technique for the orbitals involves computing
the overlap matrix between MO's of the $(m-1)$th and $m$th step
\begin{equation} \label{eq:orb_ov}
\begin{aligned}
 S^{m-1,m}_{ij}= &\langle\psi^{m-1}_{i}|\psi^{m}_{j}\rangle \\
               = &\sum_{\mu\nu}C_{i\mu}^{m-1}C^{m}_{j\nu}\langle\phi^{m-1}_{\mu}|\phi^{m}_{\nu}\rangle
 \end{aligned}
\end{equation}
where $\langle\phi^{m-1}_{\mu}|\phi^{m}_{\nu}\rangle$ are the 
AO overlap matrix elements of $(m-1)$th and $m$th steps. 
For the active space, we choose the orbitals at step $m$ with
maximum overlap with the active space orbitals at step $m-1$. 
Eq.~(\ref{eq:orb_ov}) also allows us to align the MO phases 
for adjacent geometry optimization steps. 

\subsection{Excited state tracking in DMRG}\label{subsec:maxov_mps}

We further use maximum overlap of the DMRG wavefunctions 
to target and track the correct state-specific excited state solution.
Within the standard ground-state sweep algorithm at a given geometry, 
the desired excited state can usually be found in the eigenspectrum 
at the middle of the sweep (when the renormalized Hilbert space is largest) but
can be lost at the edges of the sweep when the renormalized Hilbert space is small (if it
is generated for the incorrect eigenvector). To keep following the excited state across
the sweep by generating the appropriate renormalized Hilbert space, we 
ensure that at each block iteration we always pick the Davidson solution 
with maximum overlap with the excited state solution at 
the previous block iteration. Between geometries, we ensure that we are 
tracking the correct excited state by computing the
overlap between the DMRG wavefunctions at the different geometries.
In principle, this requires multiplying the overlaps 
between the $\mathbf{L}^\sigma$, $\mathbf{R}^\sigma$ matrices, and $\mathbf{c}$ vectors. 
However, we find it is sufficient (and of course cheaper) to only 
compute the overlap between the $\mathbf{c}$ vectors 
for the two geometries, at the middle of the sweeps.

The state-specific DMRG wavefunction maximum overlap scheme is:
 \begin{enumerate}\label{enum:maxov_mps_step}
 \item \label{enum:maxov_mps_step.1} At the initial geometry, use a state-averaged DMRG 
algorithm to obtain initial guesses for $n$ states~\cite{Dorando2007}. (The more robust two-site DMRG algorithm may be used here~\cite{White1992}, and a highly accurate initial guess 
for a small M can be obtained by running back sweeps from large M~\cite{Roberto2015}). 
Store the wavefunction vectors $\{\mathbf{c}^{i}$\} (for $i = 1,2,...,n$) at
the middle of the sweep. Note that in the state-averaged procedure 
all $n$ states share the same left and right rotation 
matrices $\{\mathbf{L}^{\sigma}\}$ and $\{\mathbf{R}^{\sigma}\}$.
 \item \label{enum:maxov_mps_step.2} At a new geometry optimization 
step (=initial geometry in the first step), restart the DMRG sweep with 
the same M from the previous solution for the targeted excited state, and 
use state-specific DMRG with the \textit{one-site} sweep algorithm to 
get the new solution for the targeted excited state. (Note that 
any noise in the DMRG algorithm should be turned off). 
At each block iteration, apply the following steps in the Davidson solver:
  \begin{enumerate}
  \item \label{enum:maxov_mps_step.2.1} Perform DMRG wavefunction 
prediction by Eq.~(\ref{eq:qr_decomp}) from the previous block iteration, to 
obtain guess vectors $\{\mathbf{c}^{i}_{guess}$\} for the current block iteration.
  \item \label{enum:maxov_mps_step.2.2} Perform the Block-Davidson algorithm to 
obtain solutions $\{\mathbf{c}^{i}_{sol}$\}.
  \item \label{enum:maxov_mps_step.2.3} Compute overlaps between 
vectors $\{\mathbf{c}^{i}_{sol}$\} and $\{\mathbf{c}^{i}_{guess}$\}, and 
align the phases when needed.
  \item \label{enum:maxov_mps_step.2.4} Choose the new 
solution $\mathbf{c}^{x}_{sol}$ in $\{\mathbf{c}^{i}_{sol}$\} for the 
targeted excited state, from the largest overlap 
between $\mathbf{c}^{x}_{sol}$ and $\mathbf{c}^{n}_{guess}$.
  \item \label{enum:maxov_mps_step.2.5} Store the vector $\mathbf{c}^{x}_{sol}$ 
as the new solution. 
 \end{enumerate}
\item \label{enum:maxov_mps_step.3} Repeat Step.~\ref{enum:maxov_mps_step.2} in 
further geometry optimization steps.
\end{enumerate}

\section{Excited state geometries of trans-polyenes}\label{sec:polyene}

Excited state geometry optimization in linear polyenes 
serves as a starting point to understand the photophysical and photochemical 
behaviour of analogous systems, such as the carotenoids, in biological processes. 
We take as our systems, 
the \textit{trans}-polyacetylenes C$_{2n}$H$_{2n+2}$, with $n = 5-10$. 
We modeled the excited states and geometry relaxation as follows:
1) We obtained ground state S$_{0}$ (1$^{1}$A$_{g}$) geometries 
with DFT/B3LYP\cite{b3lyp}. 
2) We then used the DFT ground state geometries as initial guesses 
to perform ground state geometry optimization with DMRG-CI analytic energy gradients. 
3) We recomputed excited states at the DMRG optimized ground state geometries. 
4) We then further relaxed the excited state geometries with the DMRG-CI gradients. 
All calculations were performed with 
the {cc}-pVDZ basis set\cite{ccpVDZ_3db,ccpVDZ_3da, ccpVDTZ_H_B-N}. The 
active spaces were chosen as ($n$e, $n$o), where $n$ is the total 
number of $\pi$ electrons. We identified the $\pi$ active spaces 
consisting of carbon $2p_{z}$ orbitals from the L\"{o}wdin MO population analysis 
at the initial geometry, and tracked the active spaces through the geometry relaxation 
with the orbital maximum overlap method in Sec.~\ref{subsec:maxov_mo}. 
 We also carried out additional calculations with a second ``energy-ordered'' active space,
consisting of the lowest $\pi$ and $\sigma$ orbitals to make up an ($n$e, $n$o) active space. We clearly distinguish when we are referring to the second active space
in the  discussion below.
The initial ground state DFT/B3LYP geometry optimizations were carried 
out with the \textsc{Molpro} quantum chemistry package\cite{MOLPRO}. 
State-specific DMRG wavefunctions were obtained with 
the \textsc{Block} DMRG program~\cite{Chan2002, Chan2004, Sharma2012, Website}, 
using the state-specific and adiabatic wavefunction tracking 
by wavefunction maximum overlap in Sec.~\ref{subsec:maxov_mps}. 
DMRG-CI gradients were implemented in the \textsc{ORCA} quantum chemistry package. 
All calculations worked in the canonical HF orbital basis (no localization). 
To improve the geometry optimization we employed approximate nuclear Hessians, updated by the 
BFGS method~\cite{BFGS1, BFGS2, BFGS3, BFGS4}.

To simplify the analysis, in this
work we only considered  geometry optimization {\it in the plane}. Non-planar
geometries are of course relevant to polyene excited states 
but even at planar geometries, important features of the electronic excited state geometries
(e.g. the solitonic structure) appear and remain to be understood at an {\it ab-initio} level. The planar optimization
was not enforced explicitly other than through a planar initial guess, and
otherwise the coordinates were allowed to relax in all degrees of freedom. 
Consequently, electronic wavefunctions were computed 
within \textit{C}$_{1}$ spatial point group symmetry. We used 
three different numbers of renormalized states $M$=100, 500, 1000 to obtain 
DMRG wavefunctions for all states, to examine the 
influence of wavefunction accuracy on the geometries. 
Converging DMRG wavefunctions to a high accuracy ensures
the accuracy of the particle density matrices, which then ensures that the 
correct geometric minima can be reached.
However, when the magnitude of gradients was much larger than the unconverged DMRG error, 
(for example, when the geometry was far from the equilibrium)
loose DMRG convergence and fewer sweeps were used to 
decrease the computational time.

To further characterize the low-lying excited states, we analyzed 
the exciton and bimagnon character of state transitions using the 
transition particle density matrices. 
The first-order transition density matrix element in the MO basis between 
the ground (GS) and excited states (ES) is
\begin{equation}\label{eq:single_excitation}
\left\langle \Psi_{ES} \left| {c}_{i}^{\dag} {c}_{j} \right| \Psi_{GS} \right\rangle 
\end{equation}
where $i$ and $j$ denote spatial MO indices.
We used the first-order transition density matrix to locate the first 
optically dark and bright states by the following well established state signatures:
1) A single  large element where $i$=LUMO, $j$=HOMO, indicating 
the first optically bright state. 
2) Two dominant elements where $i$ = LUMO + 1, $j$ = HOMO and $i$ = LUMO, $j$ = HOMO - 1 
indicating the first optically dark state. 
Real space particle-hole excitation patterns were further analyzed 
by the real space first-order transition density matrix, which was obtained 
by transforming the \textit{vir-occ} block of the MO first-order transition density matrix 
to the orthogonal $2p_{z}$ basis. Real space particle-hole excitation patterns were characterized by 
excitations of an electron from an orbital at $R - r/2$ to an orbital to $R + r/2$, where 
$R$ was set at the centre of a polyene chain, and $r$ is the particle-hole separation length.  
We illustrate the excitons graphically by plotting $\left \langle {c}_{p}^{\dag} {c}_{n-p} \right\rangle$, where 
$p$ is the index of the carbon $2p_{z}$ orbital, and $n$ is the total number of $2p_{z}$ orbitals in the chain.

Similarly, the real space bimagnon character is characterized by the real-space double-spin 
flip transition density
\begin{equation}
\left\langle \Psi_{ES} \left| {c}_{p,\sigma}^{\dag} {c}_{p,-\sigma} {c}_{n-p, -\sigma}^{\dag} {c}_{n-p,\sigma} \right| \Psi_{GS} \right\rangle \label{eq:dsf}
\end{equation}
where $\sigma = \{\uparrow, \downarrow\}$. The real space second-order  
transition density matrix was transformed from the \textit{vir-vir-occ-occ} block 
of the MO basis second-order transition density matrix. 

Analogously to previous studies, we further examined bond orders and 
geometrical defects (solitons) through the bond length alternation (BLA) function 
$\delta_{n}$
\begin{equation}
{\delta}_{n}={(-1)}^{n+1}({x}_{n+1}-{x}_{n})
\end{equation}
where $n = 1,...,N_{bond}$, and $x$ denotes bond lengths. 
For even-site \textit{trans}-polyacetylenes, the two edge bonds at the ground state 
are always double bonds, thus $\delta_{n}$ will always be positive. 
Consequently, negative values of $\delta_{n}$ indicate 
a reversed bond order, and a vanishing ($\delta_{n}=0$) value 
comes from two equal bond lengths, i.e., an undimerized region. 

\subsection{State signatures and geometries}

\subsubsection{Ground state S$_{\textbf{0}}$}\label{subsec:gs}
The ground state of polyenes is denoted by the symmetry label 1$^{1}$A$_{g}$ (here we are 
using symmetry labels characteristic of idealized $C_{2h}$ symmetry) and the 
relaxed ground state geometries are planar and dimerized. For the ground state, DMRG wavefunctions 
with $M$ = 100 are sufficient to achieve qualitative accuracy in bond lengths 
of our studied polyenes.  $M$ = 500 is sufficient for quantitative accuracy.
For example, $M$ = 100 produced errors of no more than 0.006 \r{A}  
for C$_{20}$H$_{22}$, while $M$ = 500 converged the bond lengths to 
an error of 0.0003 \r{A}, as compared to bond lengths using $M$ = 1000 (near exact).
This finding is consistent with the ground state wavefunction 
of even-carbon trans-polyenes being mostly a single-determinant, and thus accurately
 described by  DMRG in the canonical molecular orbital basis with small $M$.

\begin{figure}[t]
\begin{center}
   \centering
   \includegraphics[width=0.5\textwidth]{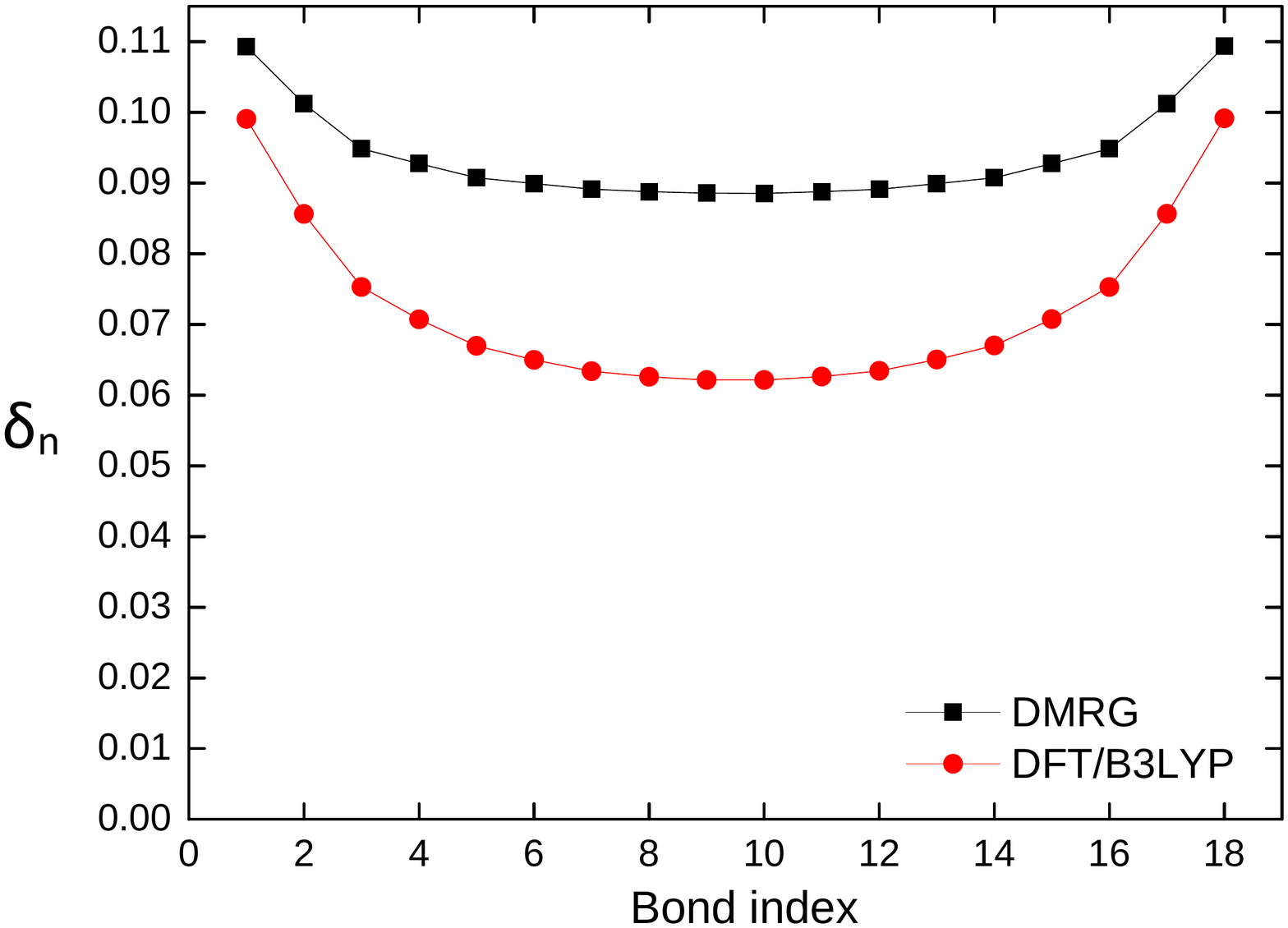}
\end{center}
\caption{Bond length alternation function $\delta_{n}$ for relaxed ground state 
geometries of C$_{20}$H$_{22}$, from left to right. }
\label{fig:BLA_S0}
\end{figure}

The BLA function $\delta_{n}$ of the 
relaxed ground state geometry of C$_{20}$H$_{22}$ from DMRG and DFT is shown in Fig.~\ref{fig:BLA_S0}.
The BLA functions from both DMRG and DFT give the same pattern, showing 
a weaker dimerization in the middle region compared to the edges of the carbon chain.
Compared to the dimerization in DFT, the dimerization in the DMRG calculations is suppressed, 
indicating a smaller dimerization gap. Compared to DFT, 
a $\pi$-active space Hamiltonian (as used in the DMRG calculations) is associated with a larger effective Coulumb interaction $U$ due to the lack of dynamic correlation. A suppression of 
the dimerization can then be expected, as 
the dimerization magnitude behaves as $U^{-3/2}$ in the
 strongly interacting limit~\cite{nakano1980solitons}.

\begin{figure}[htbp]
\begin{center}
   \centering
   \includegraphics[width=0.5\textwidth]{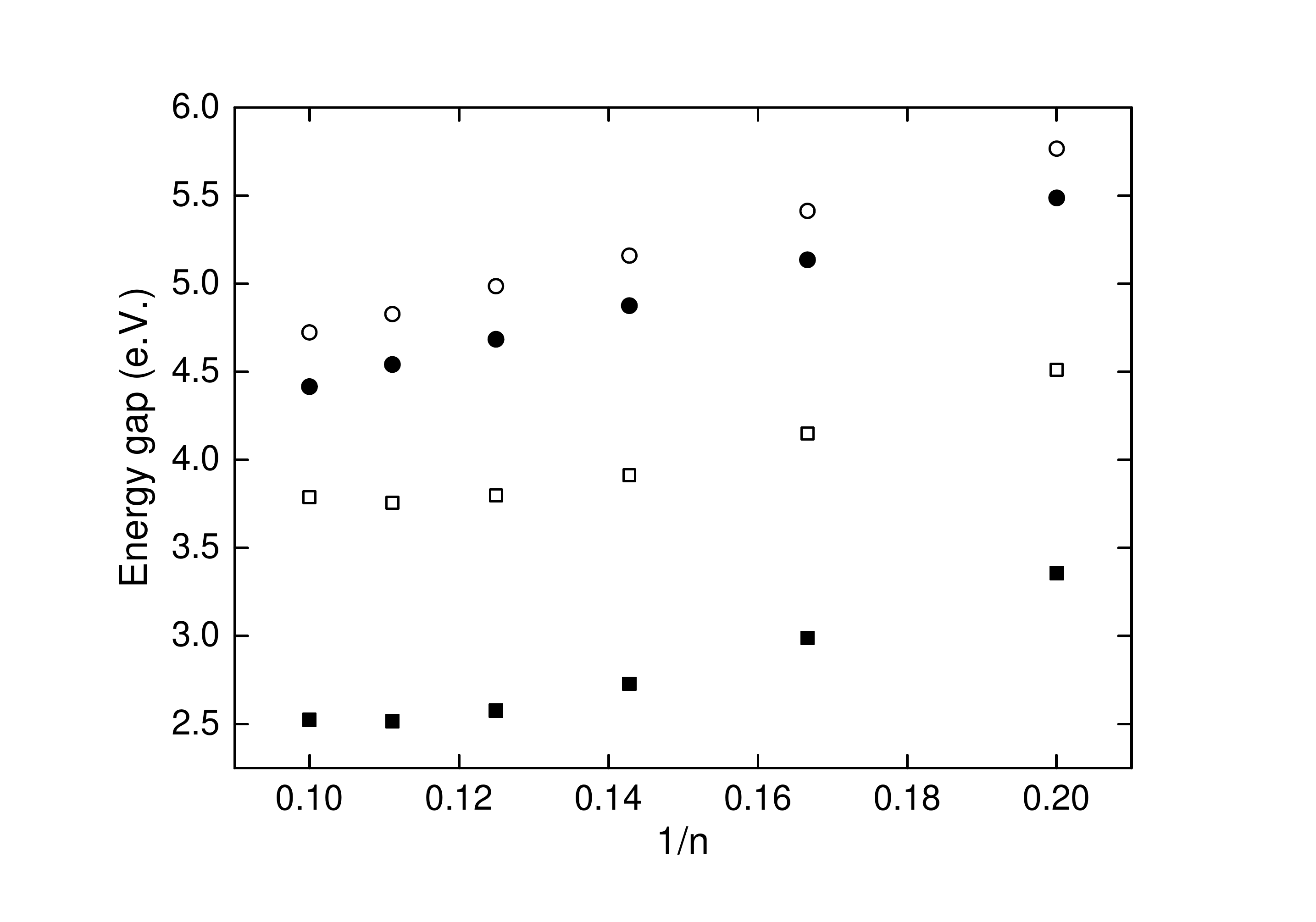}
\end{center}
\caption{Vertical and relaxed excitation energies from DMRG optimized geometries: 
vertical S$_{0}$-S$_{1}$ (opened squares), vertical S$_{0}$-S$_{3}$ (opened circles), 
relaxed S$_{0}$-S$_{1}$ (solid squares), relaxed S$_{0}$-S$_{3}$ (solid circles).}
\label{fig:Energy_gap}
\end{figure}

\subsubsection{Excited states}\label{subsec:es}
The first optically bright state in the single-$\pi$ complete active space
is the {\it third} excited state S$_{3}$, and  denoted by the 
symmetry label 2$^{1}$B$_{u}$. 
The corresponding MO based first-order transition density matrix 
between S$_{3}$ and S$_{0}$ (defined by Eq.~(\ref{eq:single_excitation})) possesses an 
element $\sim$ 1.0, where $i$ = LUMO and $j$ = HOMO, along with other elements $\le$ 0.1.
This signals a (HOMO $\rightarrow$ LUMO) single particle-hole transition,
characteristic of the first optical transition.

The 1$^{1}$B$_{u}$ state corresponds to the second
excited state S$_{2}$ in the single-$\pi$ active space. Notable first-order excitations 
in the S$_{0}$/S$_{2}$ transition, for instance in C$_{10}$H$_{12}$, 
are (HOMO $\rightarrow$ LUMO + 2) and (HOMO - 2 $\rightarrow$ LUMO) 
excitations, both with elements $\sim$ 0.5 at the ground state equilibrium geometry. 
A large (HOMO $\rightarrow$ LUMO) excitation is missing 
for the S$_{0}$/S$_{2}$ transition for all the polyenes, ruling it out as the usual
bright state.
Note that the order of excited states depends on the 
choice of  active space, i.e., the effective Hamiltonian. 
If one changes from the single-$\pi$ active space to
an energy-ordered active space which includes both $\sigma$ and $\pi$ orbitals within 
the ($n$e, $n$o) active space window, one finds that the 1$^{1}$B$_{u}$ state 
is an S$_{2}$ state corresponding to the physical optically bright HOMO $\rightarrow$ LUMO transition.
This demonstrates the well-known strong effect of dynamical correlation 
on the low-lying excited state order in linear polyenes.

The first optically dark state is the S$_{1}$ state, denoted by 2$^{1}$A$_{g}$. 
The S$_{0}$/S$_{1}$ transition exhibits dominant (HOMO $\rightarrow$ LUMO + 1) 
and (HOMO - 1 $\rightarrow$ LUMO) single excitations, along with a dominant 
(HOMO, HOMO$\rightarrow$ LUMO, LUMO) double excitation. 
The position of this low-lying excited state remains as the S$_{1}$ 
in an energy-ordered active space. 

For the bright state, optimized bond lengths were not strongly dependent on
 $M$. For a small system such as C$_{10}$H$_{12}$, $M$ = 100 produced 
a largest error of  0.0003 \r{A} in the bond lengths, as compared to the $M$ = 1000 
result. 
For a larger system such as  C$_{20}$H$_{22}$, the bond lengths at $M$ = 100 differed 
the ones at $M$ = 1000 by no more than 0.005 \r{A}, 
and the largest error at $M$ = 500 was only 0.0006 \r{A}.
For the dark state, however, the precision of the optimized geometry 
was more sensitive to the choice of $M$ for the longer polyenes. 
This may not be surprising, as the 
first optically bright state is mainly a single-reference state, while 
the lower dark state has more challenging multi-reference character~\cite{carotenoid.1}.
For all the polyenes considered, if we use small $M$, the largest error in the bond lengths of the dark state occurs for bonds around the geometrical defects (solitons).
In C$_{20}$H$_{22}$, the largest error at $M$ = 100 is about 0.025 \r{A}, 
coming from the bonds C$_{3}$-C$_{4}$ and C$_{16}$-C$_{17}$ which are around 
 the solitons (see in Sec.~\ref{subsubsec:soliton}). 
On the other hand, central bonds in the dark state are much less dependent on $M$, e.g.
$M$ = 100 yields errors $\le$ 0.012 \r{A} 
for bonds from C$_{6}$-C$_{7}$ to C$_{13}$-C$_{14}$ in C$_{20}$H$_{22}$. 
In a localized real space view, this behaviour reflects the strong
localization of multi-reference electronic structure around  the geometrical defects.

\subsection{Excitation energy}\label{subsubsec:energy}

We show vertical and relaxed excitation energies as a function of $1/n$ for 
the first optically dark (2$^{1}$A$_{g}$) and first optically bright (2$^{1}$B$_{u}$) states for all considered C$_{2n}$H$_{2n+2}$ in Fig.~\ref{fig:Energy_gap}. 
Compared to the experimental excitation energies for C$_{10}$H$_{12}$ to C$_{14}$H$_{16}$ 
in hydrocarbon solutions~\cite{kohler1988polyene}, our relaxed excitation energies are 0.3 eV higher for the  relaxed dark state, and  1.7 eV higher for the bright state.
This is in part due to the lack of dynamic correlation in our calculations as well
as basis and solvent effects.

The dark state is always observed as below the bright state.
We observe  relaxation energies for all the polyenes of about 0.35 eV 
for the bright state and about 1.20 eV for the dark state. 
The substantial relaxation energy for the dark state is consistent with the much
 larger geometry relaxation as compared to the bright state~\cite{Barford2002}.

\begin{figure}[t]
\begin{center}
   \centering
   \includegraphics[width=0.5\textwidth]{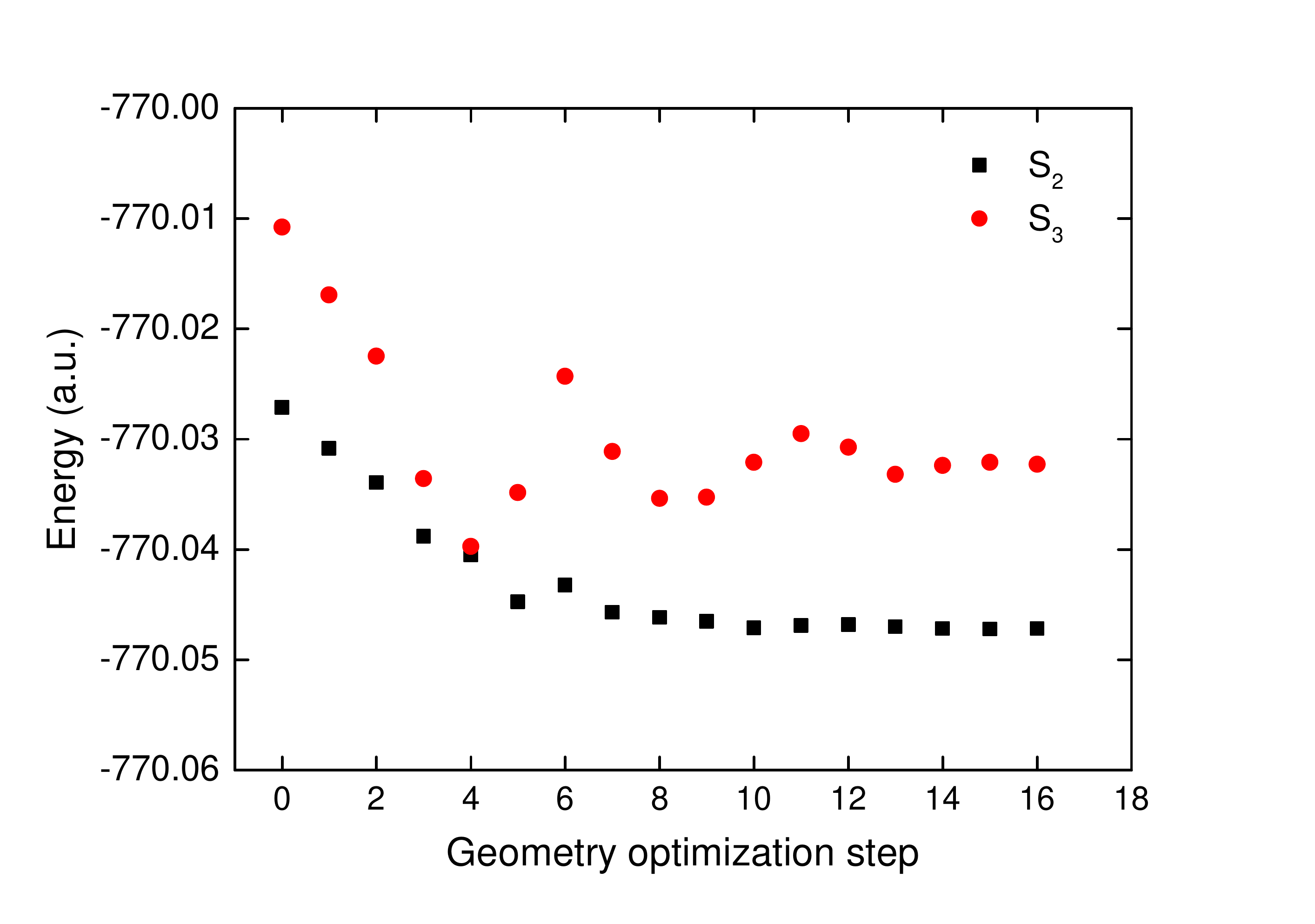}\\
\end{center}
\caption{Energies of the S$_{2}$ and S$_{3}$ states in the 
S$_{2}$ geometry optimization of C$_{20}$H$_{22}$, computed with the energy-ordered 
active space, as a function of the geometry relaxation step. At the ground state equilibrium (step 0), the S$_{2}$ and S$_{3}$ states are 
1$^{1}$B$_{u}$ (first bright state) and 2$^{1}$B$_{u}$ states respectively. At step 4 the molecule gives a S$_{2}$ and S$_{3}$ gap of 0.019 eV,
strongly indicative of a conical intersection.
After this step the 1$^{1}$B$_{u}$ and 2$^{1}$B$_{u}$ states are swapped in terms of the state energy order. (Note that the S$_3$ state
energy oscillates as only the S$_2$ state energy is being minimized). The molecular geometry remains planar along the relaxation. }
\label{fig:CI}
\end{figure}

Our calculations find the 1$^{1}$B$_{u}$ state to lie relatively close to the 
2$^{1}$B$_{u}$ state at the ground state equilibrium geometry, 
with a 1$^{1}$B$_{u}$-2$^{1}$B$_{u}$ energy gap consistently about 0.27 eV for all the polyenes. 
Given the small magnitude of this energy gap, it seems likely that there can be an energy
crossing between 2$^{1}$B$_{u}$ and 1$^{1}$B$_{u}$ states. If we use the energy ordered active space we do find an energy crossing 
between these states for C$_{20}$H$_{22}$ at a 
planar geometry, near the Franck-Condon region (Fig.~\ref{fig:CI}). Of course we also expect non-planar conical intersections, as previously found in butadiene~\cite{levine2008optimizing, Martinez2009} and octatriene~\cite{Liu2013}.

\begin{figure}[t]
\begin{center}
   \centering
   \includegraphics[width=0.5\textwidth]{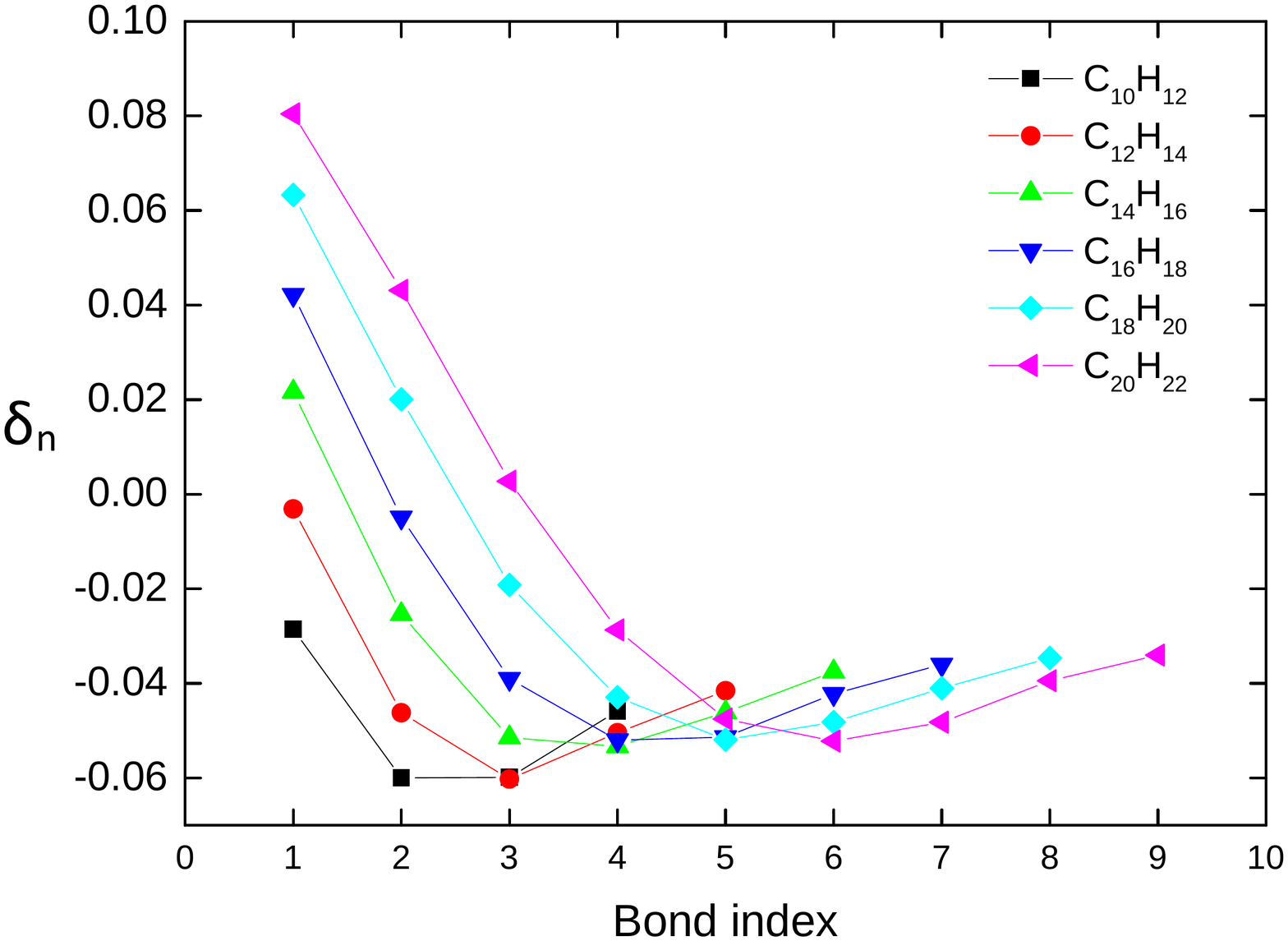}
\end{center}
\caption{Bond length alternation function $\delta_{n}$ for relaxed first dark state geometries, from edge (left) to center (right). }
\label{fig:BLA_dark}
\end{figure}

\begin{figure}[t]
\begin{center}
   \centering
   \includegraphics[width=0.5\textwidth]{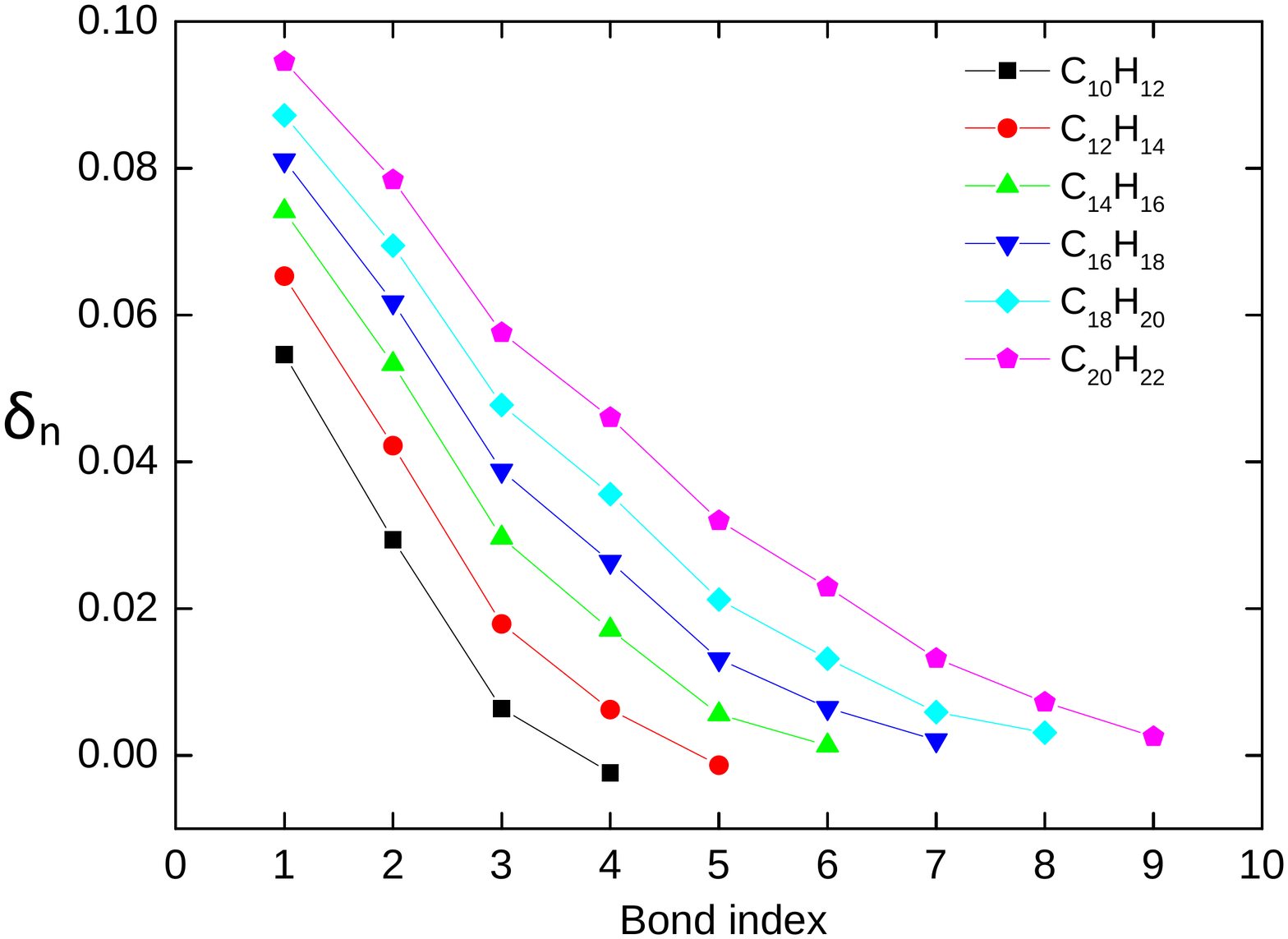}
\end{center}
\caption{Bond length alternation function $\delta_{n}$ for relaxed first bright state geometries, from edge (left) to center (right). }
\label{fig:BLA_bright}
\end{figure}

\subsection{Solitons}\label{subsubsec:soliton}

The BLA $\delta_{n}$ functions for the first optically dark (2$^{1}$A$_{g}$) and 
first optically bright (2$^{1}$B$_{u}$) states are shown 
in Figs.~\ref{fig:BLA_dark} and \ref{fig:BLA_bright}. These curves 
are almost parallel across all the polyenes for the dark and bright state respectively,
indicating generally similar behaviour across the systems.

For the 2$^{1}$A$_{g}$ state,  the BLA in short polyenes C$_{10}$H$_{12}$ 
and C$_{12}$H$_{14}$ is  completely reversed from the ground-state, as shown by the
all negative $\delta_{n}$ 
values along the chain. 
The reversal of BLA in 2$^{1}$A$_{g}$ in short polyenes has previously been understood
in terms of the dominant valence bond configurations~\cite{Hirao1996} 
  with reversed BLA.
For long polyenes, undimerization emerges near the edges 
as shown by changes in the sign of the $\delta_{n}$ functions, and 
the BLA is opposite on the two sides of the undimerized regions. 
This result is in agreement with earlier semi-empirical studies on long polyenes~\cite{Barford2001, Barford2002}, and our result shows the
two-soliton structure in the relaxed 2$^{1}$A$_{g}$ state.

For the 2$^{1}$B$_{u}$ relaxed geometry,  $\delta_{n}$ systematically shows a polaronic 
defect in the chain centre. This is also consistent 
with previous semi-empirical studies\cite{Barford2001, Barford2002}. 
For short polyenes, the vanishing dimerization in the central 
region can be understood in terms of ionic VB configurations along the chain~\cite{Hirao1996}.
In terms of excitons, the  polaronic geometry is also viewed
as evidence of a  bound particle-hole excitation localized
near the chain centre\cite{barford_book}.

\subsection{Excitons}\label{subsubsec:exciton}
\begin{figure}[t]
\begin{center}
   \centering
(a)\\
   \includegraphics[width=0.5\textwidth]{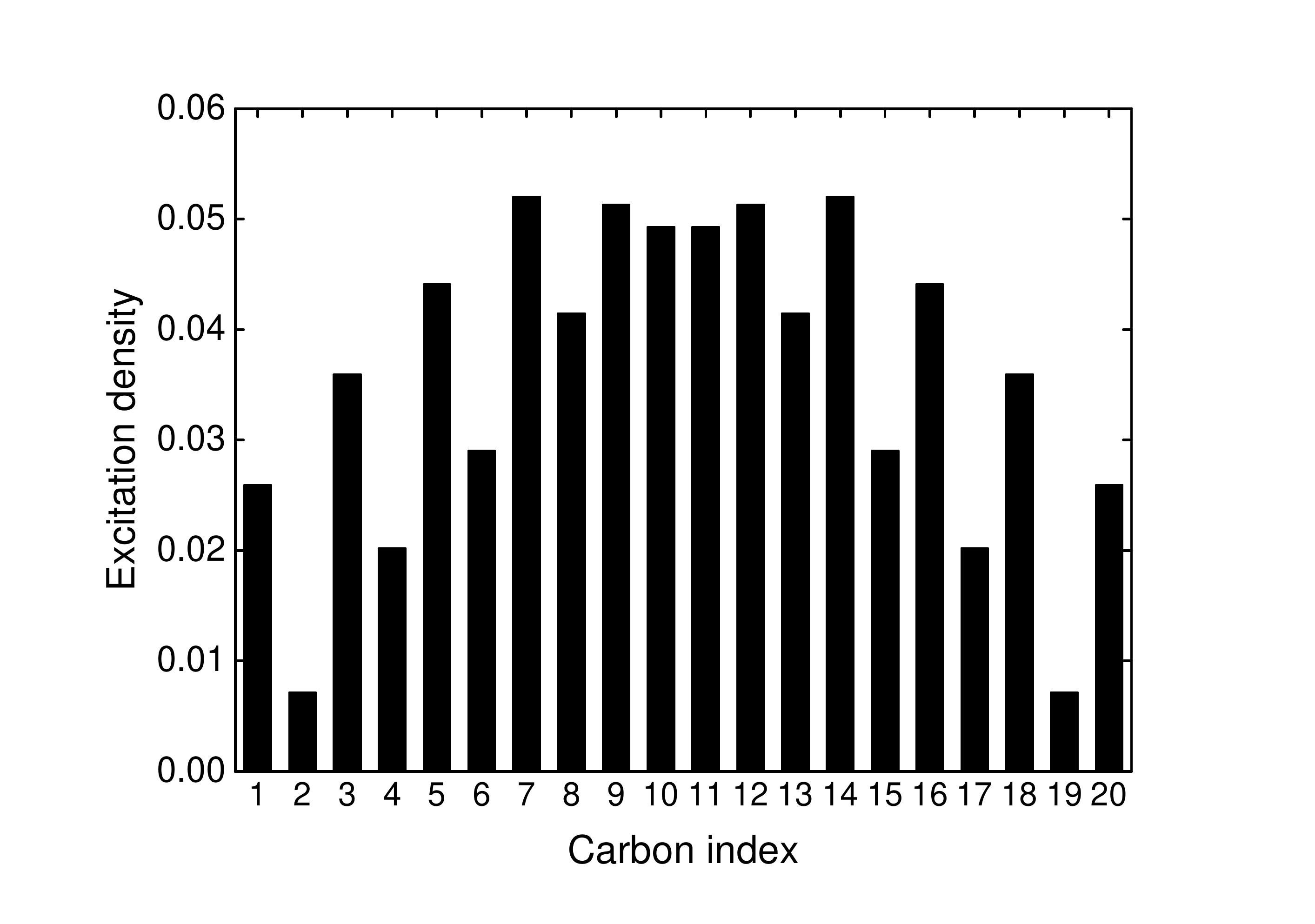}\\
(b)\\
   \includegraphics[width=0.5\textwidth]{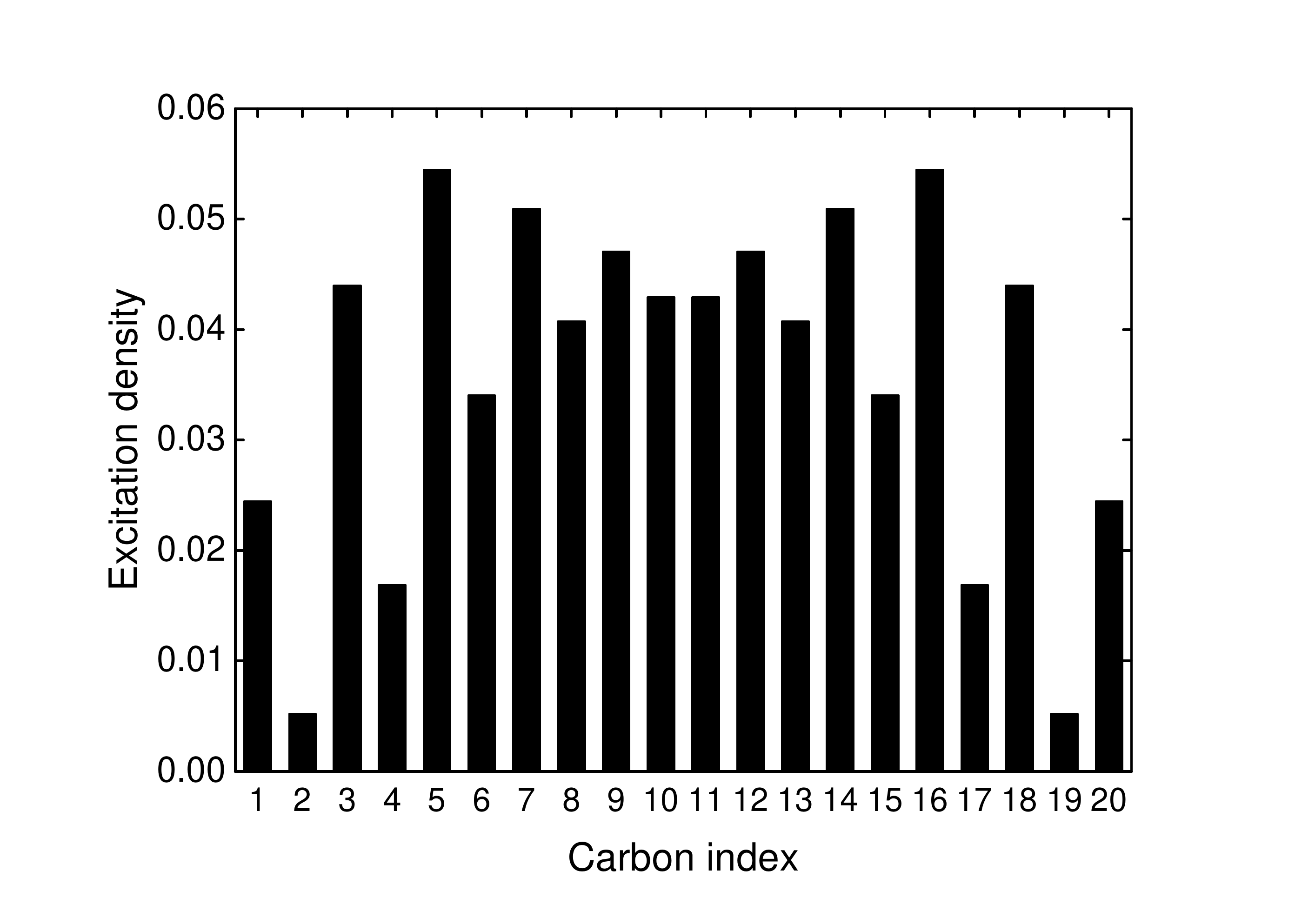}\\
\end{center}
\caption{Real space particle-hole excitation density of C$_{20}$H$_{22}$ between 
ground state and first bright state, computed at relaxed geometries 
of (a) ground state (b) first bright state.}
\label{fig:c20h22_exciton_bright}
\end{figure}

\begin{figure}[t]
\begin{center}
   \centering
(a)\\
   \includegraphics[width=0.5\textwidth]{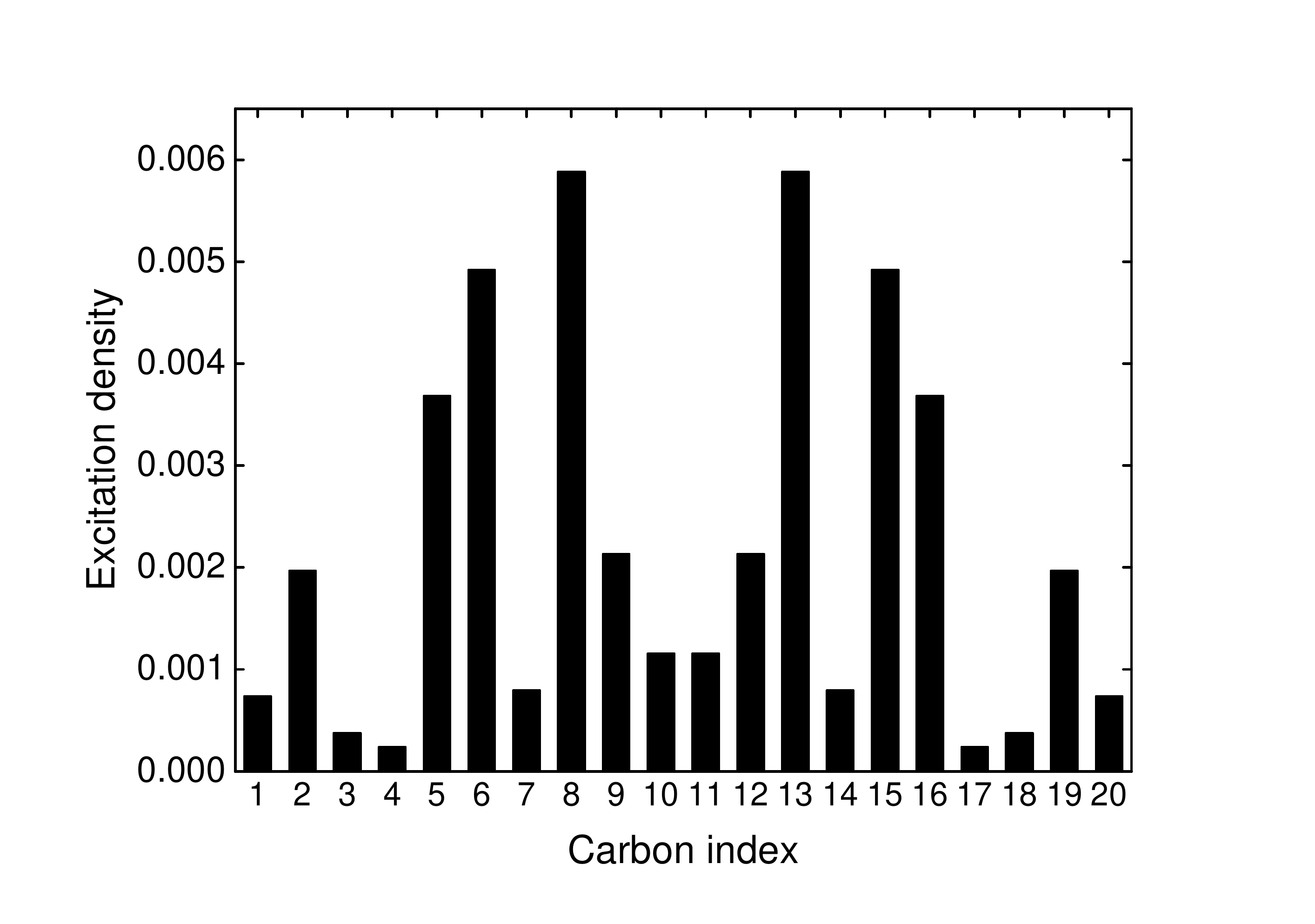}\\
(b)\\
   \includegraphics[width=0.5\textwidth]{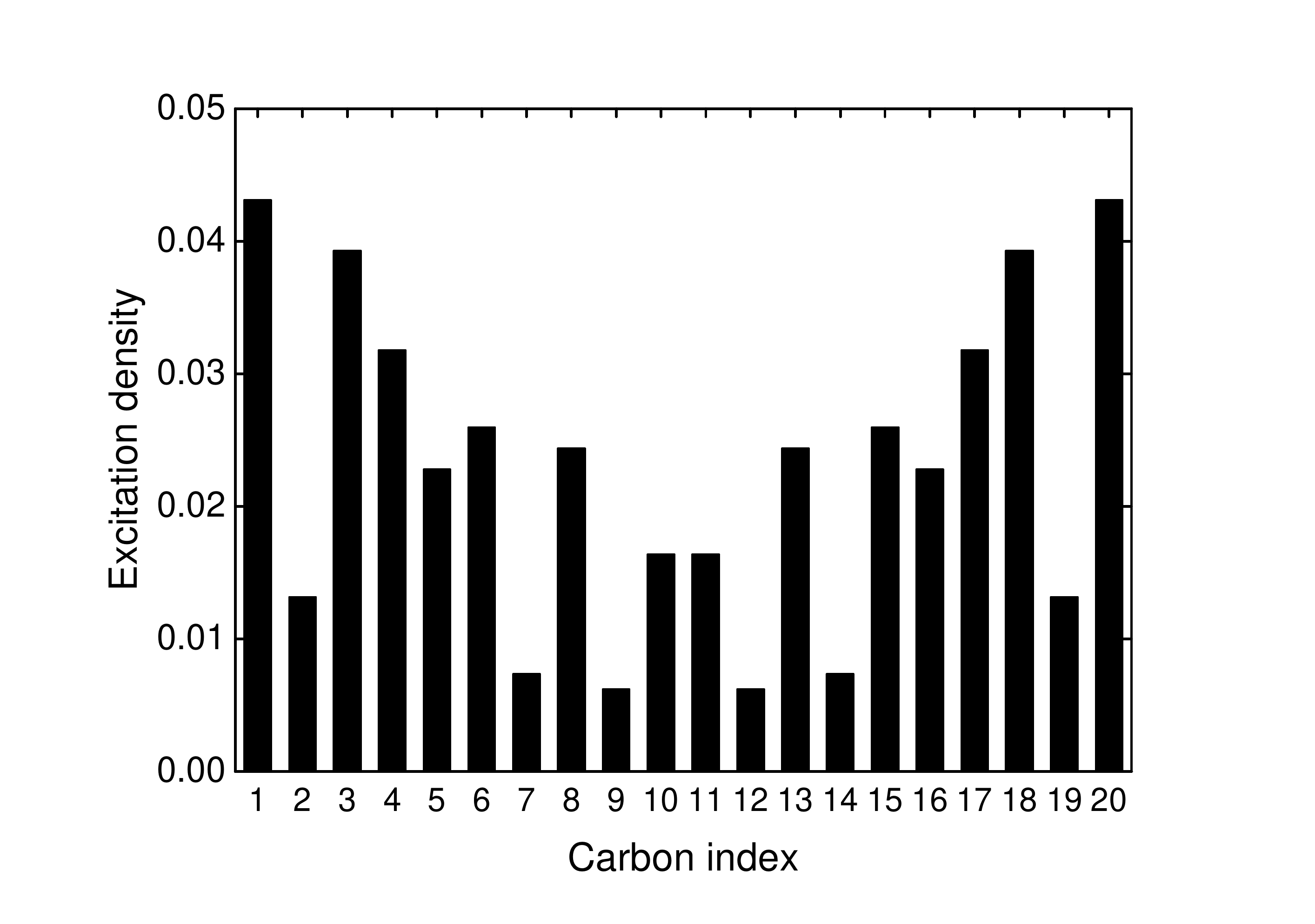}\\
\end{center}
\caption{Real space particle-hole excitation density of C$_{20}$H$_{22}$ between 
ground state and first dark state, computed at 
relaxed geometries of (a) ground state (b) first dark state.}
\label{fig:c20h22_exciton_dark}
\end{figure}

Within the one-electron manifold, we can visualize the excitons with 
the real space particle-hole 
excitation density $\left \langle {c}_{p}^{\dag} {c}_{n-p} \right\rangle$. 
As we relax the geometry, we can observe the shape of the exciton change.
Geometry relaxation is important to overcome 
the exciton self-trapping~\cite{barford_book}, e.g., in a polyene chain in 
its dimerized ground-state geometry.
The real space particle-hole excitation densities of C$_{20}$H$_{22}$ are shown 
in Fig.~\ref{fig:c20h22_exciton_bright} and Fig~\ref{fig:c20h22_exciton_dark}, 
for the bright and dark state respectively.

At the ground state equilibrium geometry, i.e., a dimerized geometry, the 
particle-hole excitations of the bright state are strongly bound, 
as seen in Fig.~\ref{fig:c20h22_exciton_bright}(a).
This is similar to as seen in the single-peak real space exciton structure 
from DFT-GWA-BSE calculations~\cite{Louie1999}, as well as the $n$ = 1 Mott-Wannier 
exciton pattern in the weak-coupling limit~\cite{Barford2001}. 
For the dark state, particle-hole pairs are slightly separated
 at the dimerized geometry, as illustrated by the double-peak 
real space exciton structure in Fig.~\ref{fig:c20h22_exciton_bright}(a). 
This has  been identified  in previous studies~\cite{Louie1999, Barford2001}, 
as an $n$ = 2 Mott-Wannier exciton.
However, the amplitudes of the densities are ten times smaller as compared to 
that of the bright state, 
essentially suggesting  neglegible exciton character for the dark state 
reached by a vertical transition. 

After geometry relaxation, the particle-hole separation in the bright state 
increases, although the  particle-hole pair remains bound 
at the bright state equilibrium geometry, as shown in Fig.~\ref{fig:c20h22_exciton_bright}(b). 
For the dark state, however,  geometry relaxation seems to unbind the 
particle-hole pair,  as shown by largely separated peaks in Fig.~\ref{fig:c20h22_exciton_dark}(b). 
Along with the enhanced transition density amplitude, this suggests the  
emergence of a long-distance charge-transfer character associated with the dark state equilibrium geometry.

\subsection{Bimagnons and singlet fission in 2$^{\textbf{1}}$A$_{\textbf{g}}$}\label{subsubsec:singlet_fission}
\begin{figure}[t]
\begin{center}
   \centering
   \includegraphics[width=0.5\textwidth]{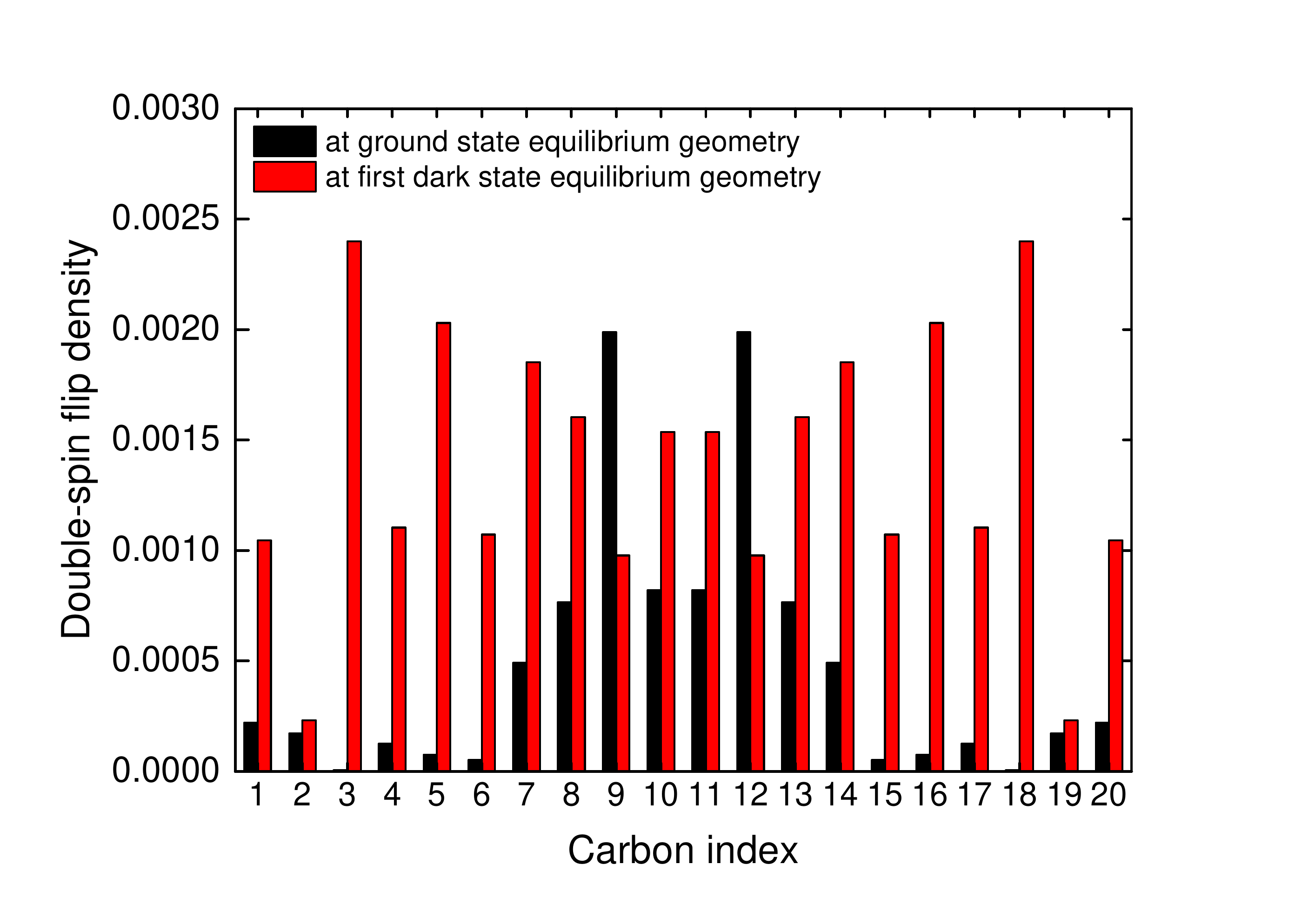}
\end{center}
\caption{Real space double-spin flip density between the ground state and the first dark state. }
\label{fig:c20h22_transition_spin_density}
\end{figure}

The relaxed dark state 2$^{1}$A$_{g}$ geometry possesses a separated two-soliton 
structure as discussed in Sec.~\ref{subsubsec:soliton}. The locally undimerized 
regions in the relaxed 2$^{1}$A$_{g}$ state can be thought to arise
from a form of ``internal singlet fission''~\cite{barford_book}, i.e., forming 
local triplets (magnons) while the total spin remains a singlet. The local triplets 
can be identifed from the local peaks of the real space spin-spin correlation function 
of the 2$^{1}$A$_{g}$ wavefunction as in Ref.~\cite{Barford2001}. 

Here, we can also characterize the bimagnon character by the real space double-spin flip  
transition density between the S$_{0}$ and 2$^{1}$A$_{g}$ states (see Eq.~(\ref{eq:dsf})). 
We show the real space double-spin flip transition density of C$_{20}$H$_{22}$ as a 
function of the site index in Fig.~\ref{fig:c20h22_transition_spin_density}. 
At the ground state equilibrium, the bimagnons are confined near the chain centre, as 
indicated by the local central double peaks. 
However, the bimagnons are highly mobile, and with geometry relaxation, 
 the singlet fission character becomes much more delocalized.
The transition density distribution possesses two peaks at carbon 3 and 17 
at the dark state equilibrium geometry, which is consistent with the positions 
of the solitons shown in Fig.~\ref{fig:BLA_dark}.

\section{Conclusions}\label{sec:conclusion}

We presented the detailed formalism for state-specific DMRG analytic 
energy gradients, including a maximum overlap algorithm that facilitates state-specific excited
state geometry optimizations. We employed these techniques to study
the ground and excited state electronic and geometric structure of the polyenes
at the level of DMRG-CI. Our quantitative results are consistent with earlier 
qualitative semi-empirical studies of the exciton, bimagnon, and soliton character
of the excited states. In addition to complex bond-length alternation patterns, 
we find evidence for a planar conical intersection.

DMRG analytic energy gradients provide a path towards the dynamical modeling 
of excited state and highly correlated quantum chemistry.
The interaction of dynamic and non-adiabatic effects with strong electron correlation 
remains an open issue, which can now be explored with the further development 
of the techniques described here.

\section{Acknowledgement}
Weifeng Hu thanks Gerald Knizia, and Bo-Xiao Zheng for discussions on the gradient theory, and thanks Sandeep Sharma for help with the \textsc{Block} DMRG code.
This work was supported by the US National Science Foundation through CHE-1265277 and
CHE-1265278.

\providecommand*\mcitethebibliography{\thebibliography}
\csname @ifundefined\endcsname{endmcitethebibliography}
  {\let\endmcitethebibliography\endthebibliography}{}

\end{document}